# Predicting final stage sintering grain growth affected by porosity

Gabriel Kerbart[1], Charles Manière[1], Christelle Harnois[1], Sylvain Marinel[1]

1 Normandie Univ, ENSICAEN, UNICAEN, CNRS, CRISMAT, 14000 Caen, France

**Abstract**

Grain growth has a definitive impact on the quality of transparent sintered materials in areas such as ballistics, biomaterials, jewelry, etc. Controlling the sintering trajectory at the precise moment of final stage sintering is one of the main sintering challenges for obtaining high-performance, fully-dense nano-ceramics. However, the final stage of sintering involves a very complex coupling between the rate of porosity elimination/grain growth and transition mechanisms. This complexity makes predicting the sintering trajectory very difficult, and most transparent material production escapes this problem by using expensive high-pressure methods such as hot isostatic pressing (HIP). In the quest for a pressureless transparent material process, this paper addresses the challenge of predicting grain growth in the transition domain from the grain growth onset (in a high porosity region) to full density for $MgAl_2O_4$ spinel. We present a comprehensive modeling approach linking theoretical models such as Zhao & Harmer's and Olevsky's equations to accurately predict the complex grain growth transition region of final stage sintering. This modeling approach opens up the possibility for numerical exploration of microstructure development via underlying kinetics experimental identification.

**Nomenclature**

$\theta$ Porosity
$\theta_c$ Critical porosity
$\dot{\theta}$ Porosity elimination rate ($s^{-1}$)
$N_g$ Mean number of pore per grain
$n, m$ Grain growth equation exponents
$\alpha$ Surface energy ($J.m^{-2}$)
$R$ Gas constant 8.314 ($J.mol^{-1}.K^{-1}$)
$T$ Temperature (K)
$\dot{G}$ Grain growth rate ($m.s^{-1}$)
$G$ Grain size (m)
$G_0$ Initial grain size (m)
$p$ Grain growth equation grain size exponent
$K$ Grain growth factor ($m^{1+p}.s^{-1}$)
$k_0$ Grain growth pre-exponential factor ($m^{1+p}.s^{-1}$)
$Q$ Grain growth activation energy ($J.mol^{-1}$)
$t$ Time (s)

*K′* Constant for isothermal condition ($m^{1+p} \cdot s^{-1}$)
*C, A* Constants
*a, b, c* fitting constants
*w* The sintering equation grain size exponent
*D* The diffusion coefficient
*k* The Boltzmann constant (1.38064852E-23 $J \cdot K^{-1}$)

### I. Introduction

The scientific community and industry takes much interest in the development of transparent polycrystalline ceramic technology due to their numerous applications for jewelry, laser hosts, spacecrafts and IR windows for military applications[1]. Ceramics' main attractive features are their thermo-mechanical properties, fracture toughness, and high hardness from room temperature to high temperatures (>1000°C). For some of them, their intrinsic transparency in the visible-IR range and their low cost of raw materials [2] also enhance their attractiveness. Several ceramic compositions present advantageous mechanical properties along with transparency, including alumina $Al_2O_3$, aluminum oxynitride spinel $Al_{23}O_{27}N_5$ and magnesium aluminate spinel $MgAl_2O_4$ [1]. Compared to alumina, which is known for its transparency and better mechanical properties, $MgAl_2O_4$ does not exhibit birefringence properties. This birefringence has an impact on transparency and implies a minimal grain size (<400 nm for alumina) in order to avoid the degradation of optical properties. Moreover, transparency can only be achieved for highly pure and fully dense materials. In the case of magnesium aluminate spinel ($MgAl_2O_4$), hereafter termed spinel, these applications also require minimal grain growth to optimize mechanical properties [3,4]. The impact of grain size on the mechanical properties of spinel is well documented. It's known that for this type of ceramic material, smaller grain size improves mechanical properties due to its Hall-Petch tendency [5].

The latest methods for sintering $MgAl_2O_4$ include Hot Isostatic Pressure (HIP) and Spark Plasma Sintering (SPS) [6–10]. These sintering methods limit grain growth through the application of mechanical pressure (gas pressure for HIP and uniaxial mechanical pressure for SPS). The best result for HIP sintering of spinel has been obtained with pretreatment, such as conventional sintering [6,7]. A pre-sintering stage followed by a HIP treatment at 1400°C leads to a transparent spinel with a grain size of 400-600 nm [6]. Similarly, Goldstein *et al* [8] obtained equivalent grain size with cold-isostatic pressing, conventional pre-sintering and HIP sintering at 1320°C-170MPa. Transparent spinel has also been obtained using SPS sintering by Bonnefont *et al* [9] who reported a grain size of 275 nm, at 1300°C-72 MPa. Sokol *et al* [10] obtained nano-sized grains of transparent spinel with a mean grain size of 50 nm by HPSPS (High Pressure Spark Plasma Sintering) at 1000MPa and 1000°C. However, those methods involve pressure and specific atmospheric conditions during sintering (argon pressure or vacuum with graphite contact, for respectively HIP and SPS). For SPS, graphite pollution has been widely reported to be a limiting transparency phenomenon [11,12]. Dopants were also studied for the sintering of spinel, notably LiF [13–15]. However, this sintering aid seems to initiate exaggerated grain growth for sintering temperature around 1620-1650°C [16].

The study's objective is to model the spinel grain growth during sintering. This modeling will be useful for predicting the sintering trajectory and therefore for finding optimized thermal treatment cycles [17]. Sintering trajectory models have proven to be a powerful tool for obtaining nano-grain, fully-dense "Zpex" grade Tosoh zirconia [18]. In this approach, the sintering response is modeled *via* a combination of densification and the grain growth model. Indeed, in addition to the detrimental effect on the ceramic's mechanical properties, grain growth slows down the sintering kinetic by extending the diffusion path [19–21]. This effect is very active in the final stage of sintering and was used to correct the sintering model densification curve at the end of sintering [22,23]. In the well-known equation (1) below, which originated from the analytical form of various solid-state pressureless sintering mechanisms [24], the final stage effect takes his origin in the term $G^w$ which divides the porosity elimination rate [19].

$$\dot{\theta} = \frac{-C\,f(\theta)\alpha D}{G^w kT} \tag{1}$$

In this formula, the grain size exponent $w$ is 3 for lattice diffusion and 4 for grain boundary diffusion [19]. Equation (1) is the base of numerous sintering methods for studying sintering mechanisms, such as master sintering curves [25], kinetic fields and the Wang and Raj approach [26], etc.

The same approach is applied to a spinel powder for controlling the sintering trajectory. We concentrate our efforts on the temperature onset region of the final sintering stage where the grain growth takes place. This transition region, where both densification and grain growth are active, represents the best opportunity for sintering optimization [27,28]. The difficulty and the interesting aspect of identifying grain growth behavior in this region lies in the influence of porosity on grain growth. At low temperatures, the grain growth kinetic is slow due to the pinning effect of the porosity located on the grain boundaries. After the separation of pores from the grain boundaries at higher temperatures, the grain growth kinetic accelerates [18]. This separation has been studied by numerous authors [29–31] and is known to be a key aspect of sintering, particularly with respect to obtaining transparent ceramics. In order to model the grain growth in this interesting transition region, we will consider the two grain growth models from Zhao and Olevsky which take into account the porosity contribution. Zhao and Harmer published a three-part study [32–34] where the grain growth is modeled theoretically via a model which takes into account porosity and pore size distribution. In the Olevsky et al model [35], porosity is implemented via a function which includes a critical porosity and easily models the transition between the region of porosity influence and the traditional form of grain growth for fully-dense materials at high temperatures.

In this work, we will compare these two models and identify the grain growth of spinel at different temperatures in the transition region where grain growth is active. Special attention will be paid to the description of the evolution of the grain growth mechanisms in the transition region.

## II. Theory and calculation

### II.1. Theoretical model of grain growth with porosity

The traditional grain growth model based on grain boundary mobility (with or without pores) obeys the following relation of the grain growth rate [19,36,37]:

$$\dot{G} = \frac{K(T)}{G^p} \tag{2}$$

with K(T) defined by various parameters that can be expressed by an Arrhenius relation:

$$K(T) = K_0 \, exp\left(\frac{-Q}{RT}\right) \tag{3}$$

For sintering, the equation (2) is often known by its integration form $G^{p+1} = G_0^{p+1} + K't$ for fully dense or porous specimens and for isothermal conditions ($K'$ constant)[38,39]. For anisothermal conditions, the term $K't$ of the integration form is replaced (mathematically) by $A \int_{t0}^{t} exp\left(\frac{-Q}{RT}\right) dt$ in order to take into account the thermal history of $K(T)$ as in the equivalent rate formulation (2) [19]. The exponent of the integrated form (*p+1* between 2 and 4) can be related to various grain growth mechanisms such as: 4 for pore control by surface diffusion, 3 for pore control by lattice diffusion or 2 for boundary control of pure system or evaporation/condensation pore drag[40]. There are numerous grain growth mechanisms and exponents but, as highlighted by Rahaman, the experimental exponent value during sintering is often close to 3 and may also evolve with temperature and additional factors such as the impurity segregation [19]. Similarly, a transition of the grain growth kinetic is often observed with temperatures corresponding to a change in the grain growth regime [18,41]. This change has supposedly been linked to a change in the pore drag regime on grain growth. In the present approach, we will investigate the impact of both the porosity and the change of the grain growth mechanism during the sintering of spinel.

The impact of the porosity on the grain growth kinetic is mainly investigated by various identified mechanisms [40] using equation (2). However, the evolution of the porosity elimination may directly impact the grain growth kinetic regardless of the exponent. Very few theoretical studies have been published in this regard [21], but one can cite the works of Zhao and Harmer [32–34]. One of their main achievements is the grain growth equation depending on the porosity $\theta$ and the mean number of pores per grains $N_g$:

$$\dot{G} = \frac{K(T)}{G^p} \left(\frac{N_g^m}{\theta^n}\right) \tag{4}$$

With p, n and m, exponents depend on the control mechanisms (Surface diffusion: *p=3, n=4/3 and m=1/3*, lattice diffusion: *p=2, n=1 and m=0*). Riedel and Svoboda propose various grain growth models with a similar form [42]. Olevsky's grain growth model corrected the traditional theory by modeling the effect of porosity on grain growth *via* a critical porosity function:

$$\dot{G} = \frac{K(T)}{G^p} \left(\frac{\theta_c}{\theta + \theta_c}\right)^n \tag{5}$$

These equations modulate grain growth by the introduction of a porosity influence function which reduces the grain growth kinetics for various porosity levels. However, when samples approach maximum density, Zhao's function of porosity tends toward infinity while Olevsky's empirical function tends toward 1, which reduces the model to its conventional form for normal grain growth in fully-dense materials.

Zhao and Harmer use equation (4) in combination with a densification model to model the sintering trajectory. However the consequence of the grain growth function singularity for full density makes the grain growth rate tend toward infinity close to full densification ($1/\theta^n \rightarrow \infty$). Furthermore, the determination of the mean number of pores per grain (*Ng*) is very difficult for classical microstructures with a high variety of pores. For this reason, Zhao and Harmer used the pore former to ease the experimental identification of *Ng*. Therefore, Olevsky's model appears to be more suitable for routine experimental use. However, it's important to understand if this empirical function describes the grain growth mechanism in the same way as the theoretical equations (like those of Zhao and Harmer) do.

In order to compare Olevsky and Zhao's and Harmer's grain growth law porosity function, the experimental results of Zhao and Harmer's function ($N_g^m/\theta^n$) were plotted in Fig. 1 for a *p* exponent of 2 and 3, corresponding respectively to surface and lattice diffusion pore control mechanisms [34]. Olevsky's function[35] in the form $a(\frac{\theta_c}{\theta+\theta_c})^n$ was also plotted in Fig. 1, with a fitting parameter "*a*" that can be assumed to be taken from the term K(T) in equation (5). It's clearly demonstrated in Fig. 1 that Olevsky's empirical critical porosity function has the same tendency of Zhao and Harmer's grain growth model. Nevertheless, compared to the latter, Olevsky's function depends only on the porosity and does not suffer from the singularity for full densification case (due to the critical porosity term). Based on these results, the following study will consider Olevsky's function to estimate the porosity influence on the grain growth of spinel. It will be demonstrated that this porosity function is inevitable for describing grain growth behavior in the transition region mentioned in the introduction.

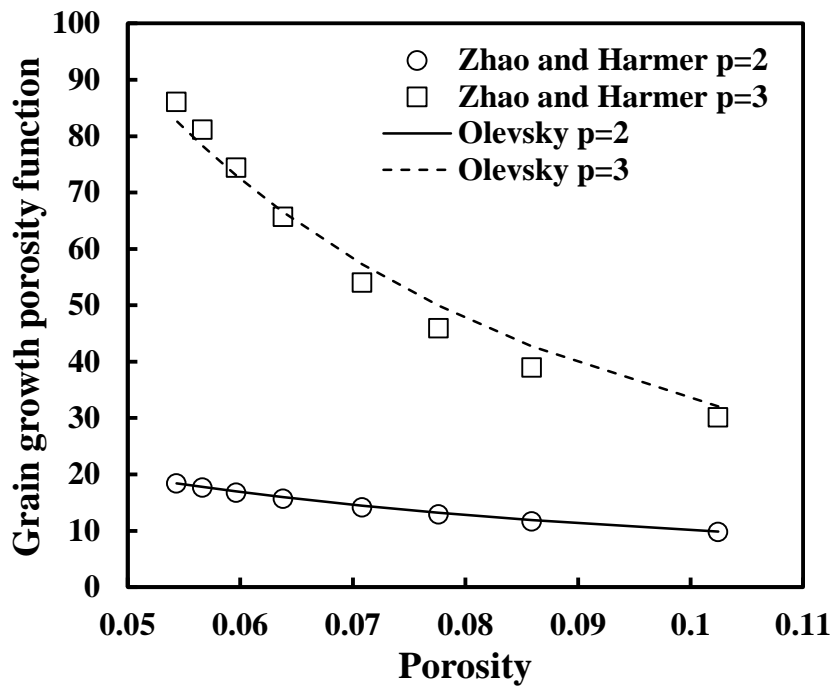

*Figure 1 Zhao and Harmer's theoretical grain growth porosity dependent term fitted by Olevsky's empirical function*

### II.2. Identification methodology

In order to identify the parameters of the grain growth model, four main steps are required and are defined below. This methodology is based on interrupted isothermal sintering tests at various temperatures and a double regression for determining the grain growth kinetics (see method scheme in Fig. 2).

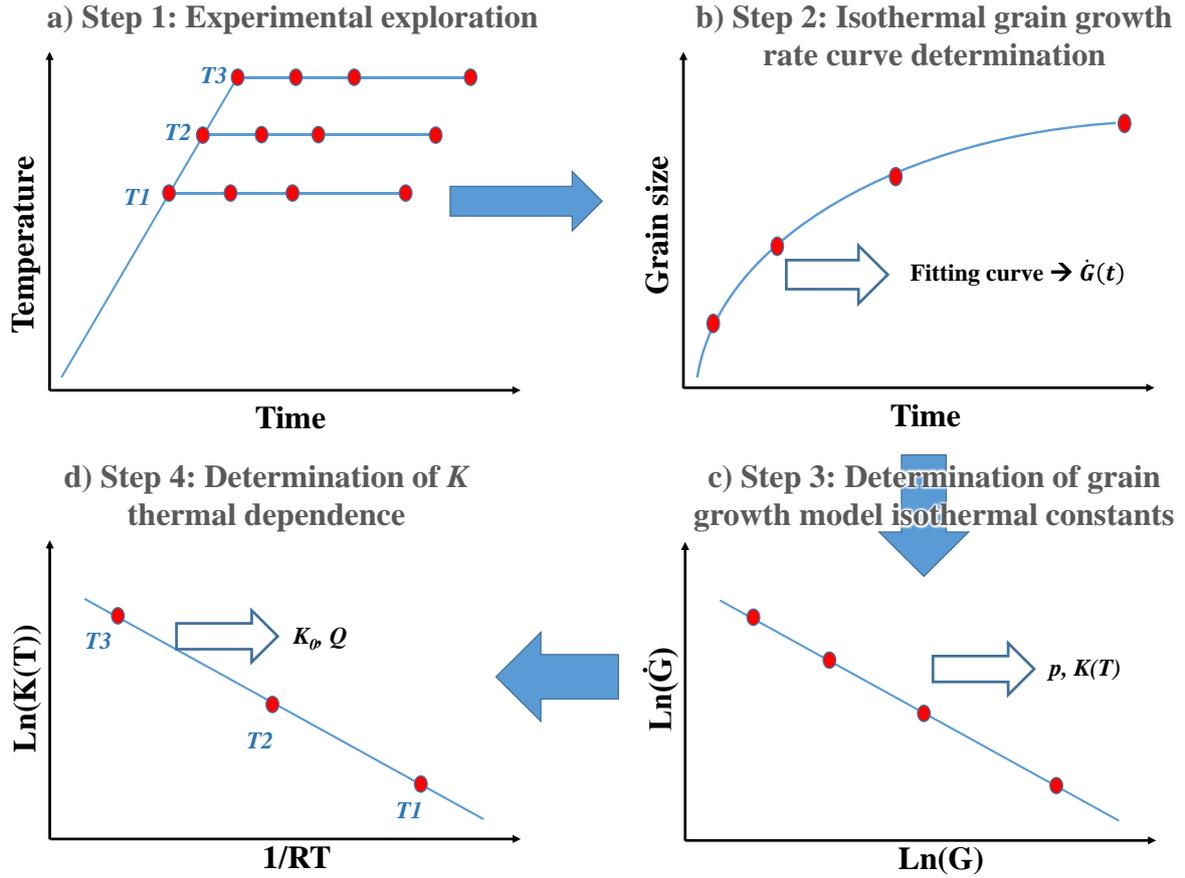

*Figure 2 The four steps of the grain growth model identification method*

Step 1 consists of the experimental collection of porosity and grain size data for obtaining isothermal grain growth curves at three different temperatures. For each temperature, more points are taken at the beginning of the dwell temperature where the grain growth is faster.

After data collection, step 2 consists of a mathematical fitting of each isothermal grain size point for determining the grain growth rate curve $\dot{G}$. The fitting mathematical equation is below:

$$G_{fit} = a + b\, ln(t + c) \tag{6}$$

At this stage, $G$ and $\dot{G}$ are known. It's then possible to determine the grain growth model constants *K* and *p* for each of the three holding temperatures (step 3). The linear regression equation appears below:

$$\ln(\dot{G}) = \ln(K(T)) - p\ln(G) \tag{7}$$

Knowing *K(T)* for the three different temperatures, the final step (step 4) is the identification of the pre-exponential constant $K_0$ and activation energy *Q* via the second linear regression below originating from equation (3):

$$\ln(K(T)) = \ln(K_0) - \frac{Q}{RT} \tag{8}$$

This traditional identification method [23,37] is used to determine the grain growth parameters. However, in this study, we chose a temperature range where the porosity has a high impact on grain growth. Consequently, the traditional method is not accurate enough to describe the results correctly (for the present study). The addition of the effect of porosity is required for a better description of the microstructure evolution. In our methodology, the porosity function is introduced during step 3 (Fig. 2c) to correct the regression curves which are not completely linear due to the porosity effect on isothermal grain growth (see later). We will also consider the grain growth exponent $p$ change with the sintering temperature.

### III. Experiment

Samples have been produced by cold isostatic pressing at 300 MPa using a commercial powder S30CR from Baikowski. This powder has a purity of more than 99%, a specific surface area between 25 and 28 m²/g (BET) and a d50 between 0.15 and 0.3 µm. The samples were sintered with dwelling times of 0min, 30min, 1h and 4h at three different temperatures with the same heating ramp (2 K/min) (see Fig. 3.a). These experiments do not readily correspond to interruptions in order to avoid the deterioration of the heating element of the furnace (cooled by inertia). Additionally, a finite element analysis of the furnace thermal inertia was conducted to estimate the error generated by the heating and cooling latency on grain growth (see the dedicated part in the results section).

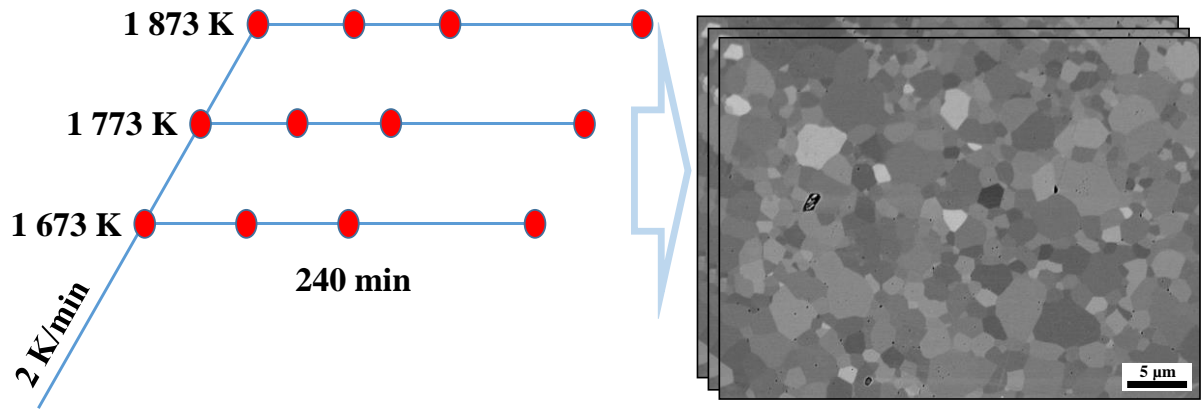

*Figure 3 a) Experimental sintering cycles. b) Microstructures of spinel polished surfaces used to determine grain size*

The mean grain size was measured using the intercept method. The measurement has been done with at least 5 lines intercepting 10 grains on different SEM images for each sample. As established in Mendelson's publication [43], a corrective factor of 1.56 has been used. The microstructures were obtained using a Jeol 7200 LV scanning electron microscope (Fig. 3.b). The micrograph acquisition was conducted under low tension of the beam (< 3keV) to avoid a charging effect on the samples' surfaces. With this method, no thermal etching was required. The analysis was conducted with BSE electrons to highlight the microstructure and grains. The sample surfaces were polished down to 1 µm with a diamond suspension and 0.25 µm with colloidal silica. Density was measured using Archimedes's method.

**IV. Results and discussion**

**IV.1. Finite element study of furnace inertia**

The furnace temperature during the sintering tests was simulated to accurately obtain the thermal history of the samples. The main purpose was to determine the potential disturbance of the furnace thermal inertia on the grain growth curves, particularly for cooling. A finite element simulation was conducted using COMSOL Multiphysics software (Fig. 4) using the same conditions of ref[44]. In order to calibrate the furnace thermal inertia of the simulation, two temperature measurements (Fig. 4. a) were made in two locations (Tfurnace & Tsample, see Fig. 4. b). Tfurnace is the temperature inside the furnace used for the PID regulation. Another thermocouple was placed close to the sample (Tsample) to detect the inertia of this zone where an alumina box was present.

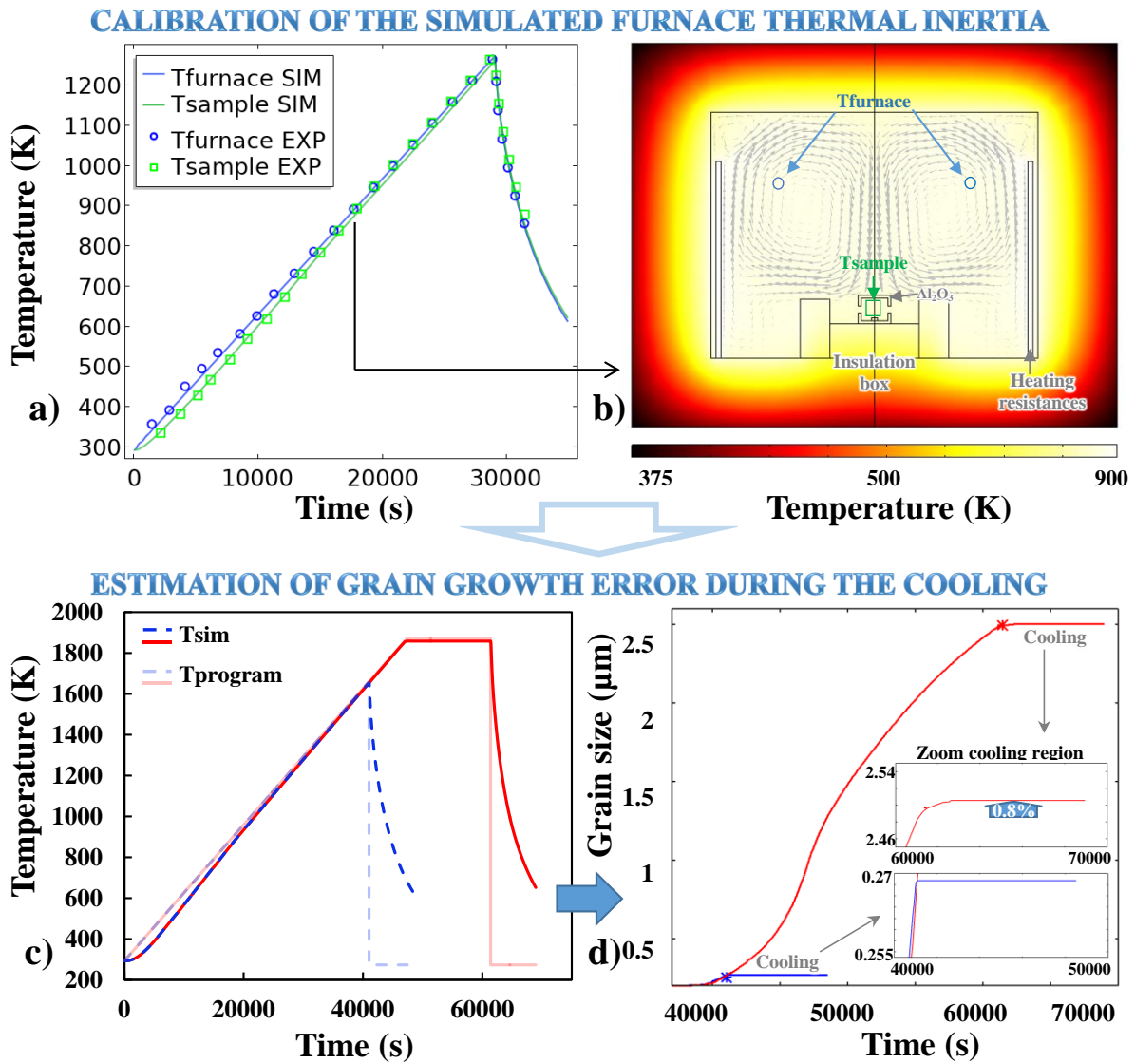

*Figure 4 a) Simulated thermal cycle compared to experimental data. b) Furnace simulation with thermocouple location. c) Difference between simulated thermal cycle and furnace programming. d) modeled grain growth for two experiments 1673 K - 0 h and 1873 K - 4 h*

We can see on Fig. 4. a) the existence of a small temperature offset during the heating ramp. This shift disappears at high temperature where sintering takes place, and therefore has no impact on the results. Furthermore, furnace cooling takes place very quickly, meaning that it may not have a significant impact on grain growth. The simulation Fig. 4. b) shows the temperature distribution inside the chamber with the convection fluxes indicated by gray arrows. To test the impact of cooling on grain growth, we modeled on Fig. 4. c) the case of the onset of grain growth 1673 K – 0 h, and the opposite case of high temperature and high density, 1873 K – 4h.

The result shown on Fig. 4. d. is obtained using the grain growth model described later in this paper. We simulated the grain growth for the sintering cycles taking into account the furnace inertia (simulated curve Fig. 4. c). We show in this figure that there is essentially no grain growth during the cooling of the 1673 K – 0 h experiment. For the 1873 K – 4h experiment, we find 0.8% grain growth (blue arrow Fig. 4. d) during cooling. These experiments, under two different conditions, show that the error on the grain growth value due to the thermal inertia is less than 1% . Therefore, this phenomenon can be ignored.

**IV.2. Step 1 dilatometry and choice of the exploration temperature domain**

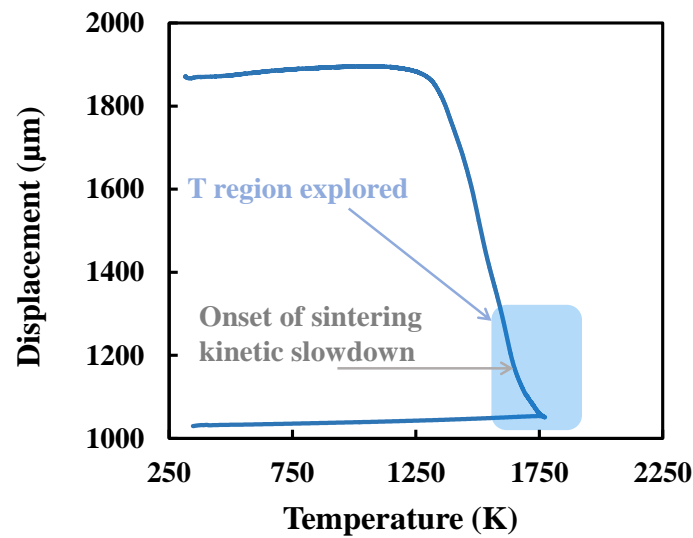

*Figure 5 Interest temperature determined by a dilatometry experiment*

A dilatometry experiment was conducted in order to identify the experimental temperatures of interest for the sintering and grain growth study (Fig. 5). On the dilatometry curve, we see a significant slowdown in densification from an onset temperature (indicated on Fig. 5), which is an indication of the grain growth onset [18,22]. Consequently, the three temperature holdings were taken in this region of interest, between 1673 K and 1873 K. Since porosity plays a major role in grain growth, long dwell times will be conducted in order to gain valuable insights into this phenomenon. Therefore, we choose to sinter the samples with dwelling times varying from 0 min to 4h.

Once determination of the region of interest is conducted, determination of grain growth parameters can be subsequently conducted. The influence of porosity on grain growth will then be deduced.

### IV.3. Identification of grain growth parameters

### IV.3.1 Step 2, collections of experimental data and mathematical fittings.

Fig. 6. a) shows the evolution of grain growth with dwell times. Equation (6) was used to determine the evolution of the grain size as a function of time for each sintering temperature. After that, the grain growth rate ($\dot{G}$) can be calculated as shown in Fig. 6. b). As expected, we notice a lower level of grain growth for lower dwell temperatures (1673 K and 1773 K) and significant grain growth at 1873 K.

From the dilatometric curve, the density evolution with the heating ramp can be obtained. With the values of the sintered samples' density, we can estimate the evolution of the porosity during the sintering process (Fig. 6. c). The analysis of the evolution of the porosity with the sintering time, correlated to the grain growth rate, seems to indicate a strong influence of porosity on the growth rate for low porosity levels.

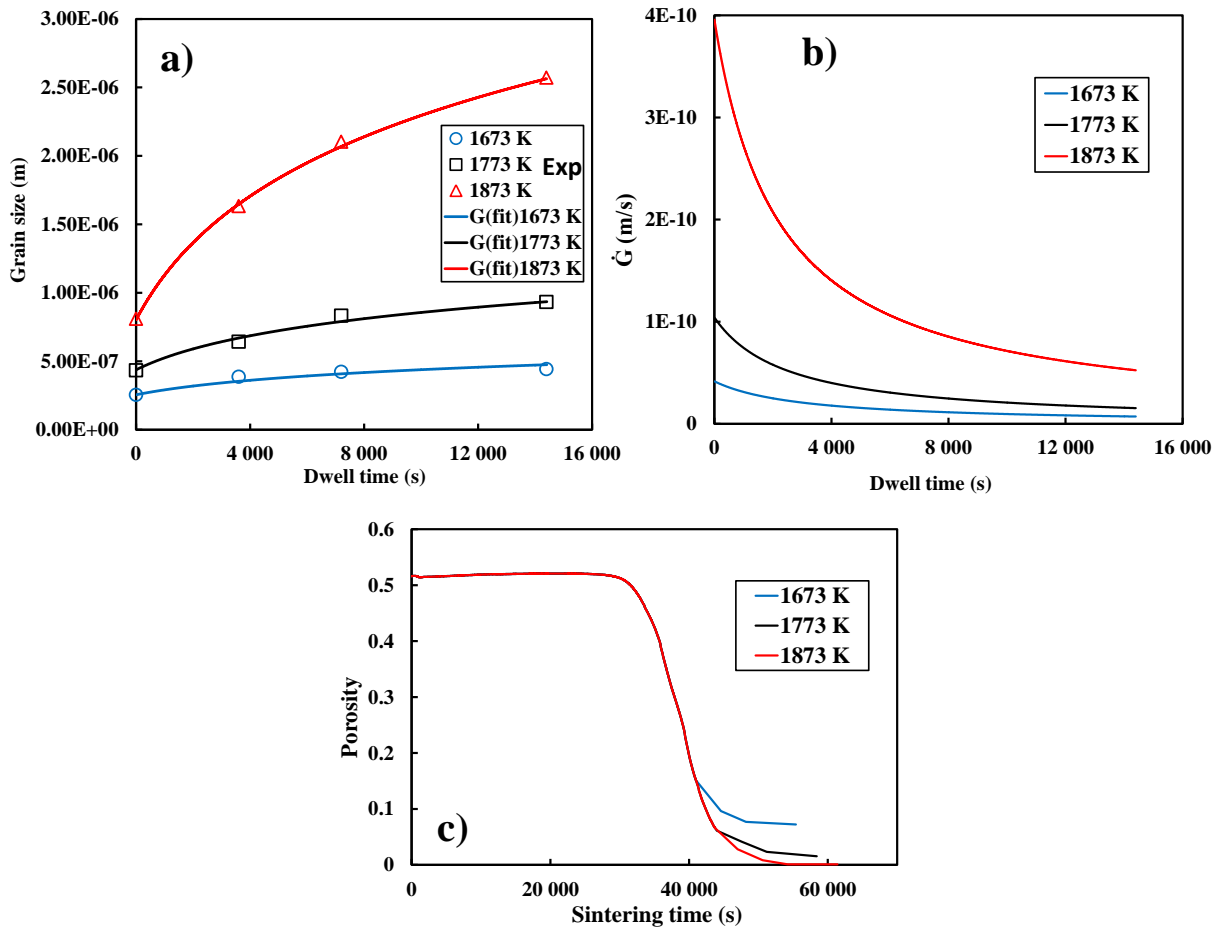

*Figure 6 a) Grain size fitted as a function of the dwell time b) grain growth rate as a function of the dwell time c) evolution of porosity during sintering*

With knowledge of the grain size and grain growth rate, we can determine the grain growth exponent p and the pre-exponential factor K(T) as stated in the equation (7).

### IV.3.2 Step 3, determination of the grain growth mechanism with porosity correction.

Fig. 7. a) presents the regression of the logarithm of grain size as a function of the logarithm of the grain growth rate. From this point onward, the identification method is split between the traditional model (the model without the porosity effect) and Olevsky's model (the model with the porosity effect). The grain growth exponent p changes from 3 to 2 with the sintering temperature. This phenomenon is apparent in both models, and can be explained by the evolution of the grain growth mechanism changing from surface diffusion for high porosity (p=3) to lattice diffusion for low porosity (p=2)[19]. Porosity diminishes with the increasing sintering temperature, which explains the transition from high porosity surface control mechanisms to lattice control.

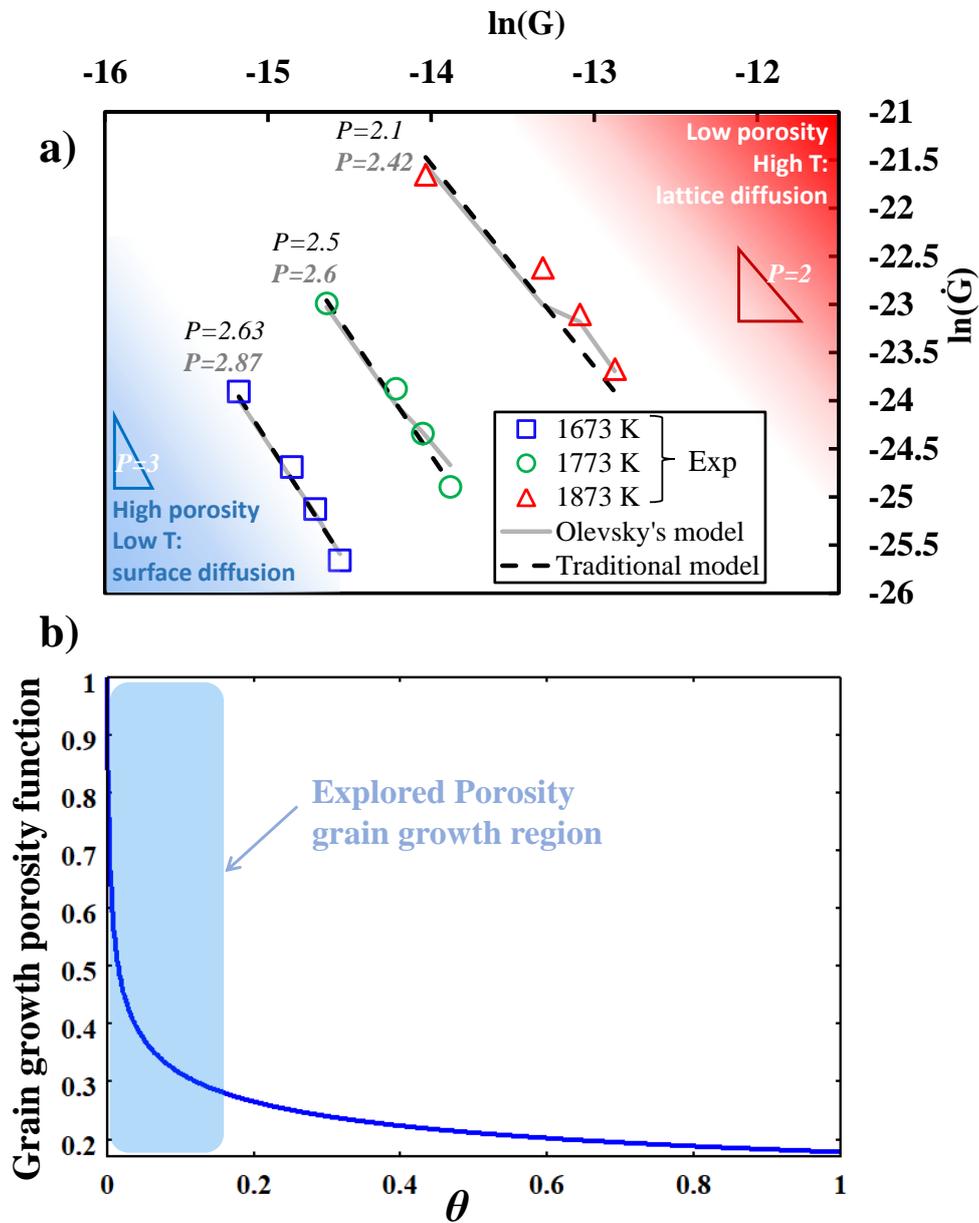

*Figure 7 a) Linear regression of the logarithm of the grain growth rate, as a function of the logarithm of the grain size. b) Olevsky's porosity function*

Olevsky's model implies a porosity function, and is shown on Fig. 7. b). The blue zone symbolizes the region of porosity we have explored in this paper. This clearly shows the presence of a 0.3 coefficient between samples with low porosity *vs* high porosity. The effect of this correction is significant for a dwelling at 1873 K, with the first points misaligned. We will see later that the porosity function also plays an important role in the modeling of the grain growth behavior versus the dwell temperatures.

With the determination of the grain growth exponent and the pre-exponential factor, the next step is to determine the activation energy of grain growth.

### IV.3.3 Step 4, determination of K0 and Q

As stated by the equation (8), the activation energy of the grain growth mechanism can be obtained by the regression of the logarithm of K(T) as a function of 1/RT. Both regressions are presented in Fig.8. The traditional model (Fig. 8. a) fits the experimental data with low accuracy, despite taking into account the change of the p exponent. The model parameters are $K_0$=1.430E24 $m^{1+p}.s^{-1}$, $Q$=1673 kJ/mol, and the high value of the activation energy is due to the change of mechanism (*p*). In comparison, Olevsky's model (Fig. 8. b) has a much more accurate alignment of the data points and a parameters value of $K_0$= 3.433E+18 $m^{1+p}.s^{-1}$, $Q$=1515 kJ/mol.

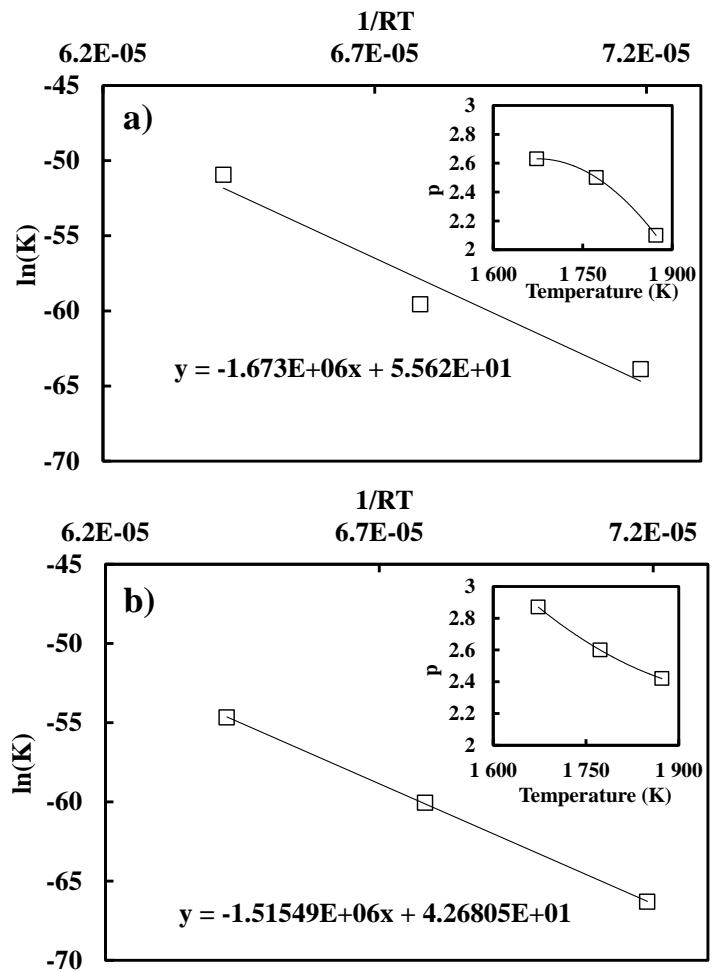

*Figure 8 a) Determining the activation energy of grain growth using traditional models (without the porosity effect), b) Determining the activation energy of grain growth using Olevsky's model*

The impact of porosity on grain growth is the only factor that can explain these differences in point alignment since it is the only change that exists between the two models. Determining the activation energy was the final step before modeling the grain growth during sintering. Two different models were devised and computed with Octave software for the traditional model (equation (2)) as well as for Olevsky's (equation (5)) model.

**IV.4 Analytical grain growth modeling without the porosity effect**

In Fig. 9., we show the results of the simulation of grain growth using the traditional model. It seems that, although the grain growth exponent change has been considered, it is not sufficient to predict the experimental data. There are two noticeable inaccuracies.

On one hand, a high level of discrepancy is noticed in the beginning of the dwell which generates a high error while the grain growth rate (slope of the curve) looks better at the end of each dwell.

On the other hand, the modeled differences between the three dwell temperatures are not well described, particularly for 1773 K. These errors are attributed to the misalignment of the data in Fig. 8. a) and to the absence of the porosity function, which has the ability to modulate grain growth with the porosity difference during each dwell and the average porosity variation between each dwell (see Fig. 6. c).

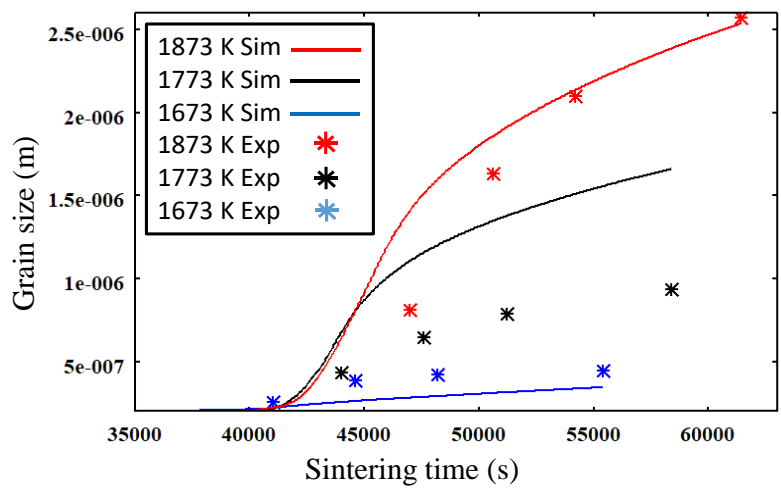

*Figure 9 Traditional grain growth model curves*

**IV.5 Grain growth modeling with the porosity effect.**

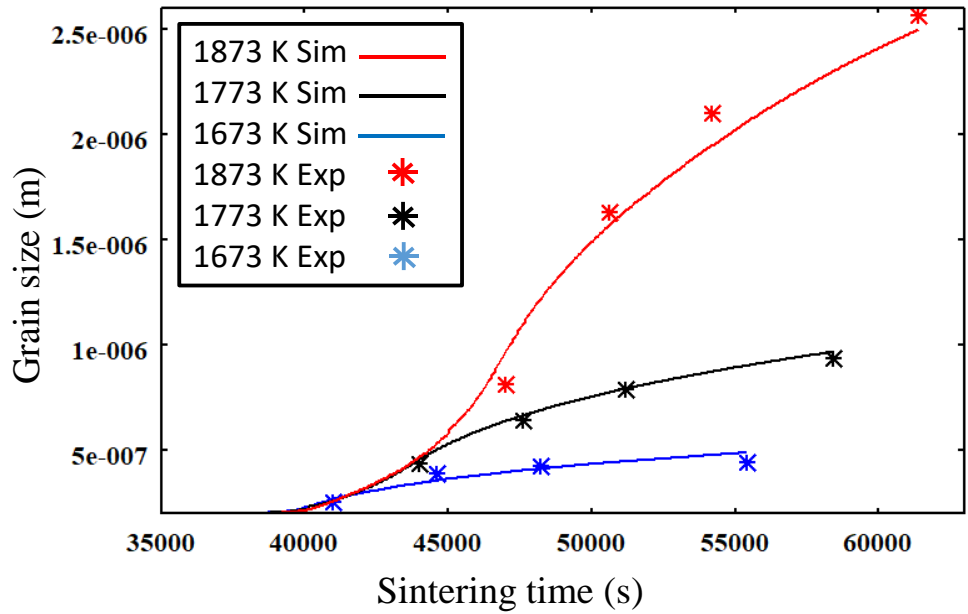

*Figure 10 Olevsky's grain growth model with the porosity effect*

In Fig. 10, we show the results of the simulation using Olevsky's model. A high correlation can be observed between the experimental data and the modeling results, especially in the early stage of grain growth. This simulation used the porosity function reported in Fig. 7. b) and the values $K_0$= 6.014E+18 m$^{1+p}$.s$^{-1}$, $Q$=1515 kJ/mol. Compared to the values from Fig. 8. b), $K_0$ was slightly adjusted to improve the overall grain growth model accuracy. In comparing Fig.9. and Fig. 10., the role of the porosity function is clearly demonstrated. This function corrects the excessive grain growth rate at the beginning of holding and very efficiently modulates the difference between the curves at various temperatures.

The corrective porosity function is then necessary to explain grain growth in the domain where a transition mechanism as well as a high porosity variation takes place. It's also important to note that the transition region, from the surface diffusion to the lattice diffusion mechanism, has a combined effect of p *vs* θ, which makes the model highly sensitive to variations compared to the traditional approach where the porosity is generally lower and is dominated by a single mechanism.

## V. Conclusion

In this paper, we explored various models of grain growth allowing for the prediction of the final stage sintering grain growth transition with porosity. The theoretical reference model for this transition region comes from Zhao and Harmer, and we can easily apply this theoretical principle to the real ceramic materials in which Olevsky's model was employed. We have shown the correlation between these two models and the limitations of the theoretical model, in particular, for the transition to full density, which is better described by Olevsky's model. In this paper, we outlined a comprehensive method based on interrupted isothermal tests to determine, step by step, the constitutive parameters of the grain growth model. Furthermore, we used a finite element simulation to estimate the influence of furnace inertia (during heating and cooling) on grain growth data collection. This simulation showed that furnace inertia may generate only a negligible deviation of the results (<1%).

We have demonstrated two main effects of porosity variation on grain growth for spinel samples:

- High porosity at low temperature favors grain growth mechanisms by diffusion at the pores' surfaces while at high temperature and lower porosity, lattice diffusion grain growth mechanisms are dominant.

- The grain growth porosity function has an important role in modulating the grain growth mechanism transition from low to high temperatures and the grain growth rate inhibition at the beginning dwell.

Olevsky's model is a relevant tool for explaining spinel grain growth in a complex but interesting range of temperatures for sintering optimization purposes. In this grain growth model, the porosity evolution was taken from the experimental curves data. However, the main endeavor of this work will be to combine this grain growth model with a sintering model in order to predict the sintering trajectory of spinel, thus avoiding a very time-consuming experimental design.

**Acknowledgements**

The author thanks Baikowski for providing us free samples of spinel powder to use in this study. The author also thanks Jérome Lecourt and Christelle Bilot for their help in the progression of this study. We would like to thank the French Ministry of Research for the PhD subvention of Gabriel Kerbart.

**Data availability**

The raw/processed data required to reproduce these findings cannot be shared at this time due to technical or time limitations.

**Credit authorship contribution statement**

**Gabriel Kerbart:** Conceptualization, Experimentation, Modeling, Writing; **Charles Manière:** Conceptualization, Supervision, Modeling, Writing; **Christelle Harnois:** Conceptualization, Supervision, Review & Editing; **Sylvain Marinel:** Conceptualization, Supervision, Review & Editing.

**References**

[1] S.F. Wang, J. Zhang, D.W. Luo, F. Gu, D.Y. Tang, Z.L. Dong, G.E.B. Tan, W.X. Que, T.S. Zhang, S. Li, L.B. Kong, Transparent ceramics: Processing, materials and applications, Prog. Solid State Chem. 41 (2013) 20–54. doi:10.1016/j.progsolidstchem.2012.12.002.

[2] J. Petit, L. Lallemant, Drying step optimization to obtain large-size transparent magnesium-aluminate spinel samples, in: B.J. Zelinski (Ed.), 2017: p. 101790K. doi:10.1117/12.2262427.

[3] J.A. Salem, Transparent Armor Ceramics as Spacecraft Windows, J. Am. Ceram. Soc. 96 (2013) 281–289. doi:10.1111/jace.12089.

[4] M. Rubat du Merac, H.-J. Kleebe, M.M. Müller, I.E. Reimanis, Fifty Years of Research and Development Coming to Fruition; Unraveling the Complex Interactions during Processing of Transparent Magnesium Aluminate ($MgAl_2O_4$) Spinel, J. Am. Ceram. Soc. 96 (2013) 3341–3365. doi:10.1111/jace.12637.

[5] M. Sokol, M. Halabi, Y. Mordekovitz, S. Kalabukhov, S. Hayun, N. Frage, An inverse Hall-Petch relation in nanocrystalline $MgAl_2O_4$ spinel consolidated by high pressure spark plasma sintering (HPSPS), Scr. Mater. 139 (2017) 159–161. doi:10.1016/j.scriptamat.2017.06.049.

[6] A. Krell, J. Klimke, T. Hutzler, Advanced spinel and sub-µm $Al_2O_3$ for transparent armour applications, J. Eur. Ceram. Soc. 29 (2009) 275–281. doi:10.1016/j.jeurceramsoc.2008.03.024.

[7] A.C. Sutorik, G. Gilde, J.J. Swab, C. Cooper, R. Gamble, E. Shanholtz, The Production of Transparent $MgAl_2O_4$ Ceramic Using Calcined Powder Mixtures of $Mg(OH)_2$ and $\gamma$-$Al_2O_3$ or AlOOH, Int. J. Appl. Ceram. Technol. 9 (2012) 575–587. doi:10.1111/j.1744-7402.2011.02679.x.

[8] A. Goldstein, A. Goldenberg, M. Hefetz, Transparent polycrystalline $MgAl_2O_4$ spinel with submicron grains, by low temperature sintering, J. Ceram. Soc. Japan. 117 (2009) 1281–1283. doi:10.2109/jcersj2.117.1281.

[9] G. Bonnefont, G. Fantozzi, S. Trombert, L. Bonneau, Fine-grained transparent $MgAl_2O_4$ spinel obtained by spark plasma sintering of commercially available nanopowders, Ceram. Int. 38 (2012) 131–140. doi:10.1016/j.ceramint.2011.06.045.

[10] M. Sokol, M. Halabi, S. Kalabukhov, N. Frage, Nano-structured $MgAl_2O_4$ spinel consolidated by high pressure spark plasma sintering (HPSPS), J. Eur. Ceram. Soc. 37 (2017) 755–762. doi:10.1016/j.jeurceramsoc.2016.09.037.

[11] K. Morita, B.-N. Kim, H. Yoshida, K. Hiraga, Y. Sakka, Assessment of carbon contamination in $MgAl_2O_4$ spinel during spark-plasma-sintering (SPS) processing, J. Ceram. Soc. Japan. 123 (2015) 983–988. doi:10.2109/jcersj2.123.983.

[12] H. Hammoud, V. Garnier, G. Fantozzi, E. Lachaud, S. Tadier, Mechanism of Carbon Contamination in Transparent $MgAl_2O_4$ and $Y_3Al_5O_{12}$ Ceramics Sintered by Spark Plasma Sintering, Ceramics. 2 (2019) 612–619. doi:10.3390/ceramics2040048.

[13] G.R. Villalobos, J.S. Sanghera, I.D. Aggarwal, Degradation of Magnesium Aluminum Spinel by Lithium Fluoride Sintering Aid, J. Am. Ceram. Soc. 88 (2005) 1321–1322. doi:10.1111/j.1551-2916.2005.00209.x.

[14] K. Rozenburg, I.E. Reimanis, H.-J. Kleebe, R.L. Cook, Chemical Interaction Between LiF and $MgAl_2O_4$ Spinel During Sintering, J. Am. Ceram. Soc. 90 (2007) 2038–2042. doi:10.1111/j.1551-2916.2007.01723.x.

[15] K. Rozenburg, I.E. Reimanis, H.-J. Kleebe, R.L. Cook, Sintering Kinetics of a $MgAl_2O_4$ Spinel

Doped with LiF, J. Am. Ceram. Soc. 91 (2008) 444–450. doi:10.1111/j.1551-2916.2007.02185.x.

[16] G. Gilde, P. Patel, P. Patterson, D. Blodgett, D. Duncan, D. Hahn, Evaluation of Hot Pressing and Hot Isostastic Pressing Parameters on the Optical Properties of Spinel, J. Am. Ceram. Soc. 88 (2005) 2747–2751. doi:10.1111/j.1551-2916.2005.00527.x.

[17] I.-W. Chen, X.-H. Wang, Sintering dense nanocrystalline ceramics without final-stage grain growth, Nature. 404 (2000) 168–171. doi:10.1038/35004548.

[18] C. Manière, G. Lee, J. McKittrick, S. Chan, E.A. Olevsky, Modeling zirconia sintering trajectory for obtaining translucent submicronic ceramics for dental implant applications, Acta Mater. 188 (2020) 101–107. doi:10.1016/j.actamat.2020.01.061.

[19] M.N. Rahaman, Sintering of Ceramics, CRC Press, 2007.

[20] R.M. German, Sintering Theory and Practice, Wiley, Wiley, 1996. http://www.wiley.com/WileyCDA/WileyTitle/productCd-047105786X.html.

[21] R.K. Bordia, S.-J.L. Kang, E.A. Olevsky, Current understanding and future research directions at the onset of the next century of sintering science and technology, J. Am. Ceram. Soc. 100 (2017) 2314–2352. doi:10.1111/jace.14919.

[22] C. Manière, L. Durand, A. Weibel, C. Estournès, Spark-plasma-sintering and finite element method: From the identification of the sintering parameters of a submicronic α-alumina powder to the development of complex shapes, Acta Mater. 102 (2016) 169–175. doi:10.1016/j.actamat.2015.09.003.

[23] C. Manière, L. Durand, A. Weibel, C. Estournès, A predictive model to reflect the final stage of spark plasma sintering of submicronic α-alumina, Ceram. Int. 42 (2016) 9274–9277. doi:10.1016/j.ceramint.2016.02.048.

[24] J.D. Hansen, R.P. Rusin, M.-H. Teng, D.L. Johnson, Combined-Stage Sintering Model, J. Am. Ceram. Soc. 75 (1992) 1129–1135. doi:10.1111/j.1151-2916.1992.tb05549.x.

[25] H. Su, D.L. Johnson, Master Sintering Curve: A Practical Approach to Sintering, J. Am. Ceram. Soc. 79 (1996) 3211–3217. doi:10.1111/j.1151-2916.1996.tb08097.x.

[26] J. Wang, R. Raj, Estimate of the Activation Energies for Boundary Diffusion from Rate-Controlled Sintering of Pure Alumina, and Alumina Doped with Zirconia or Titania, J. Am. Ceram. Soc. 73 (1990) 1172–1175. doi:10.1111/j.1151-2916.1990.tb05175.x.

[27] C.P. CAMERON, R. RAJ, Grain-Growth Transition During Sintering of Colloidally Prepared Alumina Powder Compacts, J. Am. Ceram. Soc. 71 (1988) 1031–1035. doi:10.1111/j.1151-2916.1988.tb05787.x.

[28] R. Marder, R. Chaim, C. Estournès, Grain growth stagnation in fully dense nanocrystalline Y2O3 by spark plasma sintering, Mater. Sci. Eng. A. 527 (2010) 1577–1585. doi:10.1016/j.msea.2009.11.009.

[29] C.H. Hsueh, A.G. Evans, R.L. Coble, Microstructure development during final/intermediate stage sintering—I. Pore/grain boundary separation, Acta Metall. 30 (1982) 1269–1279. doi:10.1016/0001-6160(82)90145-6.

[30] M.A. Spears, A.G. Evans, Microstructure development during final/ intermediate stage sintering—II. Grain and pore coarsening, Acta Metall. 30 (1982) 1281–1289. doi:10.1016/0001-6160(82)90146-8.

[31] R.J. BROOK, Pore-Grain Boundary Interactions and Grain Growth, J. Am. Ceram. Soc. 52 (1969) 56–57. doi:10.1111/j.1151-2916.1969.tb12664.x.

[32] J. Zhao, M.P. Harmer, Effect of Pore Distribution on Microstructure Development: I, Matrix Pores, J. Am. Ceram. Soc. 71 (1988) 113–120. doi:10.1111/j.1151-2916.1988.tb05826.x.

[33] J. Zhao, M.P. Harmer, Effect of Pore Distribution on Microstructure Development: II, First- and Second-Generation Pores, J. Am. Ceram. Soc. 71 (1988) 530–539. doi:10.1111/j.1151-2916.1988.tb05916.x.

[34] J. Zhao, M.P. Harmer, Effect of Pore Distribution on Microstructure Development: III, Model Experiments, J. Am. Ceram. Soc. 75 (1992) 830–843. doi:10.1111/j.1151-2916.1992.tb04148.x.

[35] E.A. Olevsky, C. Garcia-Cardona, W.L. Bradbury, C.D. Haines, D.G. Martin, D. Kapoor,

Fundamental Aspects of Spark Plasma Sintering: II. Finite Element Analysis of Scalability, J. Am. Ceram. Soc. 95 (2012) 2414–2422. doi:10.1111/j.1551-2916.2012.05096.x.

[36] S.J. Park, P. Suri, E. Olevsky, R.M. German, Master Sintering Curve Formulated from Constitutive Models, J. Am. Ceram. Soc. 92 (2009) 1410–1413. doi:10.1111/j.1551-2916.2009.02983.x.

[37] J. Besson, M. Abouaf, Grain growth enhancement in alumina during hot isostatic pressing, Acta Metall. Mater. 39 (1991) 2225–2234. doi:10.1016/0956-7151(91)90004-K.

[38] W.D. Kingery, B. Francis, Grain Growth in Porous Compacts, J. Am. Ceram. Soc. 48 (1965) 546–547. doi:10.1111/j.1151-2916.1965.tb14665.x.

[39] F.A. Nichols, Theory of Grain Growth in Porous Compacts, J. Appl. Phys. 37 (1966) 4599–4602. doi:10.1063/1.1708102.

[40] R.J. Brook, Controlled Grain Growth, in: Ceram. Fabr. Process. Treatise Mater. Sci. Technol., révisée, Elsevier, 2016: pp. 331–364.

[41] R. Chaim, Activation energy and grain growth in nanocrystalline Y-TZP ceramics, Mater. Sci. Eng. A. 486 (2008) 439–446. doi:10.1016/j.msea.2007.09.022.

[42] H. Riedel, J. Svoboda, A theoretical study of grain growth in porous solids during sintering, Acta Metall. Mater. 41 (1993) 1929–1936. doi:10.1016/0956-7151(93)90212-B.

[43] M.I. Mendelson, Average Grain Size in Polycrystalline Ceramics, J. Am. Ceram. Soc. 52 (1969) 443–446. doi:10.1111/j.1151-2916.1969.tb11975.x.

[44] C. Manière, T. Zahrah, E.A. Olevsky, Fluid dynamics thermo-mechanical simulation of sintering: Uniformity of temperature and density distributions, Appl. Therm. Eng. 123 (2017) 603–613. doi:10.1016/j.applthermaleng.2017.05.116.

**Figures captions**

Figure 1 Zhao and Harmer's theoretical grain growth porosity dependent term fitted by Olevsky's empirical function

Figure 2 The four steps of the grain growth model identification method

Figure 3 a) Experimental sintering cycles. b) Microstructures of spinel polished surfaces used to determine grain size

Figure 4 a) Simulated thermal cycle compared to experimental data. b) Furnace simulation with thermocouple location. c) Difference between simulated thermal cycle and furnace programming. d) modeled grain growth for two experiments 1673 K -0 h and 1873 K - 4 h

Figure 5 Interest temperature determined by a dilatometry experiment

Figure 6 a) Grain size fitted as a function of the dwell time b) grain growth rate as a function of the dwell time c) evolution of porosity during sintering

Figure 7 a) Linear regression of the logarithm of the grain growth rate, as a function of the logarithm of the grain size. b) Olevsky's porosity function

Figure 8 a) Determining the activation energy of grain growth using traditional models (without the porosity effect), b) Determining the activation energy of grain growth using Olevsky's model

Figure 9 Traditional grain growth model curves

Figure 10 Olevsky's grain growth model with the porosity effect